\documentclass[  showkeys,superscriptaddress]{revtex4-2}

\usepackage{graphics,epsfig}
 
\usepackage{graphicx}
\usepackage{dcolumn}
\usepackage{amsmath}
 
\usepackage{hyperref}
\begin{document}
\title{Quasinormal Modes and Phase Structure of Regular $AdS$ Einstein-Gauss-Bonnet Black Holes}
  \author{Yerlan Myrzakulov}
   \email{ymyrzakulov@gmail.com}
     \affiliation{Department of General \& Theoretical Physics, L. N. Gumilyov Eurasian National University,  Astana, 010008, Kazakhstan}
       \affiliation{Ratbay Myrzakulov Eurasian International Centre for Theoretical Physics
Astana 010009, Kazakhstan}
     \author{ Kairat Myrzakulov}
       \email{krmyrzakulov@gmail.com }
  \affiliation{Department of General \& Theoretical Physics, L. N. Gumilyov Eurasian National University,  Astana, 010008, Kazakhstan}
  
 \author{Sudhaker Upadhyay\footnote{Corresponding author}\footnote{Visiting Associate, IUCAA  
Pune, Maharashtra 411007, India}}
\email{sudhakerupadhyay@gmail.com}
\affiliation{Department of Physics, K. L. S. College, Magadh University,\\ Nawada 805110, India}
 
\affiliation{School of Physics, Damghan University, P.O. Box 3671641167,\\ Damghan, Iran}

\author{Dharm Veer Singh\footnote{Visiting Associate, IUCAA  
Pune, Maharashtra 411007, India}}
\email{veerdsingh@gmail.com}
\affiliation{Department of Physics, Institute of Applied Sciences and Humanities, GLA 
University, Mathura 281406, Uttar Pradesh, India.}
\begin{abstract}
In this paper, we present an exact regular black hole solution in Einstein-Gauss-Bonnet 
   coupled with nonlinear matter fields.  It is a   generalization of a regular 
Einstein-Gauss-Bonnet  black hole in  $5D$ $AdS$ spacetime.  The causal structure of the obtained solution identifies with  Boulware-Deser black hole solution, except for the curvature singularity at the 
center. It incorporates the Boulware-Deser black holes in the absence of deviation parameters. We also study the thermodynamic properties of the solution that satisfies a modified first law of thermodynamics.  Furthermore, 
we discuss the stability of the obtained black hole solution and, in this regard,  a double phase transition occurs.  Within context, we find that phase transition exists
at the point where the heat capacity diverges and, incidentally, the temperature 
attains the maximum value. We discuss the fluid nature  of the black hole  also
exhibiting critical points. The quasinormal modes of the black hole solution and their dependencies on Gauss-Bonnet coupling and deviation parameters are also 
analysed in terms of null geodesics. 
\end{abstract} 
\maketitle
\section{Introduction}\label{sec1}
The higher-order  curvature theories are useful to explore various (conceptual) 
concerns of gravity. Lovelock's theory of gravity \cite{lav,lav1,lav2} is one of such 
higher-order  curvature gravity that generalizes the  general relativity (GR) 
to higher spacetime dimensions.  In Ref. \cite{lav},  Lovelock proposed that    GR in 
$4D$ with a cosmological constant  
possessing diffeomorphism invariance, metricity and second-order
equations of motion is a unique theory of pure gravity.  The Einstein-Gauss-Bonnet 
(EGB) gravity is a particular class of Lovelock gravity  that characterizes 
 non-trivial dynamics for the higher-dimensional  gravity   having  second-order field 
equations. However,  EGB gravity in $4D$ is a topological theory that, in general, 
does not
contribute to gravitational dynamics. In the recent past, a lot of efforts have been 
made to study the  EGB gravity in $4D$. For instance, 
Glavan and Lin  proposed  the $4D$   GB theory
  by re-scaling the GB coupling constant. But the theory in    $D=4$ limit is either 
  plagued   
  by the partial breaking of diffeomorphism or by additional gravitational degrees of 
  freedom. The generalization of EGB gravity as F(GB) gravity in $4D$ was presented in  Refs \cite{Nojiri:2005vv,Nojiri:2010wj}. In another effort,  EGB gravity in $4D$ is constructed that has only two 
  dynamical degrees of freedom but it breaks the temporal diffeomorphism \cite{ao,ao1}.
  
 Black holes are one of the most fascinating objects and the subjects of active research.
 Black holes are either singular or regular  solutions of the Einstein equation of GR.  
A singular solution for the rotating counterpart of  a higher-derivative theory in Lee-Wick prescription   is explored recently \cite{dvs}.  
  The first  spherically symmetric regular black hole solution  was given by Bardeen
 \cite{bar}  
that does not hold a strong energy condition. In the recent past, people are paying much attention to  the regular black hole solutions \cite{rg1,rg2,rg3,rg4,rg5,rg6,rg7,rg8}.
An exact regular black hole solution for the EGB  coupled with non-Abelian gauge field in 4D $AdS$ spacetime  and their thermal properties are explored recently \cite{ds1}.
Recently, a $4D$ $AdS$ EGB black hole solution   with nonlinear electrodynamics (NLED) is studied \cite{Singh:2022dth}. In another recent work, the EGB massive black hole solution in $4D$ $AdS$  is 
studied also \cite{ds,Cvetic:2001bk,Nojiri:2001ae,Nojiri:2002qn}.

The  NLED  in the context of black hole physics  is a more relevant and quite suitable alternative for Maxwell (linear) electrodynamics as we know that   the real electromagnetic field remains no longer linear  at high energy  due to the influence of other physical fields.  Since  the original consideration of nonlinear electrodynamics  by Born and Infeld \cite{bi}, extensive progresses on the subject has been made \cite{nl1,nl2,nl3,nl4,nl5}.  Some  NLED  coupled to GR  may provide a better explanation for  the inflation of the universe \cite{nl6,nl7,nl8,nl9}.
Black holes with NLED are quite relevant in astrophysical observations \cite{nl10,nl11}.

{  The advantage of the study of NLED field coupled with the gravity, we obtained the regular black hole solution in the particular limits it correctly retrieves the Reissner-Nordstrom black hole. Another major difference is in the strong-field limits of Einstein’s gravity, where the exponential mass function leads to a Minkowski-flat core around, which is in striking contrary with other regular black holes \cite{Ghosh:2018bxg,Singh:2022xgi} that generally have de-Sitter core \cite{Ghosh:2020tgy,Singh:2019wpu,Kumar:2018vsm}. Therefore, the curvature of the geometry has some maximal peak in between spatial infinity and core. Although, in the $4D$ EGB gravity, all regular black holes have flat core around them \cite{Nojiri:2005vv,Nojiri:2010wj}. Therefore, this novel regular black hole share many features with other regular black holes, but there are also significant differences.}

The $AdS/CFT$ correspondence \cite{mal,mal1,wit} provides a duality between  strong  interacting quantum field theory   and weakly interacting gravity. This  is also known as the holographic duality or the gauge/gravity duality.  GB term in such correspondence may play an important role, see Refs. \cite{Nojiri:1999ji, Nojiri:2001aj}. 
  Originally, this was explored in the context of string theory but further   extended to wide domains, such as the  coupling dynamics of QCD and the electroweak theories, black holes  physics,  quantum gravity, condensed matter physics, etc. In the context of  black holes, Witten 
found that black hole  thermodynamics    in AdS spaces can get a resemblance   with the thermodynamics of  dual CFT at the high temperature \cite{wit1}.

The concept of  black hole thermodynamics originated by Bekenstein \cite{11,bak} and Hawking \cite{bak0} who realized that entropy is somehow connected to the area of the Black hole horizon.  
To a certain extent, it is clear that   entropy of the black holes is  proportional to the area of horizon \cite{str,ast,car,sol}. 
This subject was studied further extensively \cite{sud2, sud3,jy,bss,sud1, behn}
In the context of black hole thermodynamics (mechanics), it is found that the black hole system satisfies the first law of thermodynamics. The stability of dS and Nariai black hole in higher derivative gravity is discussed in Ref. \cite{Nojiri:2001ds}. Here, it is found that for certain regime   Nariai black hole is stable and does not decay into pure de Sitter space. The this connection of higher derivative gravity,  negative (or zero) Schwarzschild-(Anti)-de Sitter black entropy is found which depends on the parameters of higher derivative terms \cite{Cvetic:2001bk}.

Quasinormal modes (QNMs)  have been found an active and wide area of research \cite{q1,q2,q3,q4}. 
QNMs  are found very useful to predict the stability of the 
perturbed  black holes.   Abbott et al. (LIGO scientific collaboration and Virgo 
collaboration) detected  transient gravitational waves   \cite{lig}. 
The images of Event Horizon Telescope \cite{tel} display a prominent ring   consistent 
with the size and shape of the lensed photon shadow  of a supermassive black hole. 
These studies hint about the correspondence between QNMs and black hole shadow radius.
The correspondence 
between QNMs and shadow radius may provide a new viewpoint for the
gravitational waves which are   massless particles moving along an outmost
 unstable orbit of null geodesics. Recently, the shadow cast of the charged
  Reissner-Nordstr\"om   AdS black hole in both plasma and non-plasma medium is studied \cite{sur}.

The rest of the  sections are organized as follows. In Sec. \ref{sec2}, we consider a EGB 
gravity coupled to the NLED in  $5D$ $AdS$ spacetime and obtain a new  black hole 
solution.
We discuss the horizon structure of this new black hole solution in $AdS$ spacetime. 
  The thermodynamics  of this black hole along with stability and phase transition  
  are discussed in section \ref{sec3}. The behavior of black holes as the Van der Waals 
  fluid is reported in section \ref{sec4}. We have calculated the critical values of 
  pressure, temperature, and horizon radius and their dependencies on various 
  parameters. The QNMs for the black hole solution are calculated in section 
  \ref{sec5}.
  Finally, we summarize the results and make  final remarks in the last section.

\section{Action,  Black Hole Solutions and Horizon Structure}\label{sec2}
For the present study, we are interested in the solution 
of $5D$ EGB gravity coupled to the NLED in $AdS$ space.
The  action describing $5D$ EGB gravity coupled to the NLED in $AdS$  spacetime is 
written as  \cite{hy}
\begin{eqnarray}
S =\frac{1}{2 }\int d^{5}x\sqrt{-g}\left[  {R}
-2\Lambda +\alpha (R_{\mu \nu \gamma \delta }R^{\mu \nu \gamma \delta}-4R_{\mu \nu }
R^{\mu \nu }+R^{2}) -4P \frac{\partial {\cal H}}{\partial P}+2{\cal{H}} \right],
\label{action}
\end{eqnarray}
where ${R}$,  $R_{\mu\nu}$   and $R_{\mu\nu\lambda \sigma}$ are the Ricci scalar, Ricci 
tensor and Riemann tensor, respectively. However, $\Lambda $ and  $\alpha $  are   the 
cosmological constant related to $AdS$ length $l$ via relation $-3/l^2$ and   the 
Gauss-Bonnet coupling constant, respectively.  ${\cal{H}}(P) $ is the structure-
function  that depends
on the invariant  $P =\frac{1}{4}P_{\mu\nu}P^{\mu\nu}$ of the tensor $P_{\mu\nu}$ which 
corresponds to electric induction.
The expression for the NLED  structure function ${\cal H}(P)$  is given by
\begin{equation}
{\cal{H}}(P)= 3Pe^{-\frac{q}{M}(-2qP)^{1/3}},
\label{matter}
\end{equation}
where $q$ and $M$ are the free parameters associated with the charge and mass, 
respectively.  In the weak field limit ($P<<1$),  the   NLED  structure function 
(\ref{matter}) corresponds to the  linear electrodynamics, i.e. $ {\cal{H}}(P) \approx 
P$. The requirements to satisfy the weak energy condition  are ${\cal H}<0$ and $ \frac{\partial {\cal H}}{\partial P}>0$ 
\cite{AGB1,lbev,Balart:2014cga}. 

The field equations corresponding to the action (\ref{action}) for the metric tensor 
($g_{\mu\nu}$) and electromagnetic potential ($A_{\mu}$) are
\begin{eqnarray}
G_{\mu\nu}+H_{\mu\nu}+\Lambda g_{\mu\nu}&=&2\left(\frac{\partial {\cal H}}{\partial P} 
P_{\mu\lambda}P^{\lambda}_{\nu}-2P \frac{\partial {\cal H}}{\partial P}+{\cal{H}} 
\right),\label{en} \\
 \nabla_{\mu}P^{\mu\nu}&=&0,
\label{efe}
\end{eqnarray}
 where $G_{ab}$ and $H_{ab} $ are, respectively
\begin{eqnarray}
G_{\mu\nu}&=&R_{\mu\nu}-\frac{1}{2}g_{\mu\nu}R,\\
H_{\mu \nu }& =&-\frac{\alpha }{2}\left[ 8R^{\rho \sigma }R_{\mu \rho \nu
\sigma }-4R_{\mu }^{\rho \sigma \lambda }R_{\nu \rho \sigma \lambda
}-4RR_{\mu \nu }+8R_{\mu \lambda }R_{\nu }^{\lambda }\right.   \notag \\
&& +\left. g_{\mu \nu }\left( R_{\mu \nu \gamma \delta }R^{\mu \nu \gamma
\delta }-4R_{\mu \nu }R^{\mu \nu }+R^{2}\right) \right],
\end{eqnarray}
 \noindent Now, we are interested to obtain a  $5D$ EGB black hole solution  in  the 
 presence of NED. For this, we first write the static spherically symmetric metric as 
 follows: \begin{equation}
ds^2=-f(r)dt^2 +\frac{1}{f(r)}dr^2+r^2(d\theta^2+\sin^2\theta  d\phi^2+
\sin^2\theta\sin^2\phi \,d\psi^2 ),
\label{m1}
\end{equation}
where  $f(r)$ is the metric function which will be determined later.

{  We use the following ansatz for the antisymmetric field:
\begin{equation}
P_{\mu\nu}=2\delta^{\theta}_{[\mu}\delta^{\phi}_{\nu]}D(r)\sin^2\theta\sin\phi,
\end{equation} 
which, upon integration (\ref{efe}), eventually leads to 
\begin{eqnarray}
  P = \frac{q^2}{2r^6}.
\end{eqnarray}}
Here, we chose the integration constant as $q$.

With this antisymmetric field $P_{\mu\nu}$ and invariant $P$, the non-vanishing 
component of  Einstein field  equation  (\ref{en}) results
\begin{eqnarray}
(4\alpha f' -2 r)(f -1)-r^3f' { -\Lambda r^2}=\frac{2Mk}{r^3}e^{-k/r^2},
\label{eom1}
\end{eqnarray}
where the prime ($'$)  is the   derivative of the metric 
function  $f(r)$ concerning $r$ and deviation parameter $k=q^2/M$. The solution  of Eq. 
(\ref{eom1}) determines the form of metric function as
\begin{equation}
 f(r)=1+\frac{r^2}{4\alpha}\left(1\pm\sqrt{1+\frac{8 M \alpha }{r^4}e^{-k/r^2}{ +} 
 \frac{{ 8}\Lambda\alpha}{3}}\,\right).
\label{eqnf}
\end{equation}
 {We note that the solution (\ref{sol1}) has two branches, $+ ve$ and $- ve$, respectively.  

For vanishing Mass, the obtained black hole solution (\ref{sol1})  becomes
\begin{equation}
f(r)=1+\frac{r^2}{4\alpha} \left(1\pm \sqrt{1-\frac{8\alpha}{l^2}}\right).
\label{sol1}
\end{equation}
For $\alpha > 0$,  $8\alpha/l^2\leq 1$ and beyond this, there is no black hole solution. Thus, the action \ref{action} has two $AdS$ solutions with effective cosmological constants $l^2_{eff}=\frac{l^2}{4} \left(1\pm \sqrt{1-\frac{8\alpha}{l^2}}\right)$.  For $8\alpha/l^2 = 1$, both the solutions coincide and, therefore, the theory has a unique $AdS$ vacuum.

When $\alpha < 0$ , the solution (\ref{sol1}) still  remains $AdS$ for $- ve$ signature and becomes $dS$ if one takes the $+ ve$ signature.  From the vacuum case, the solution (\ref{sol1}) with both signs seems reasonable, from which we cannot determine which sign  should be adopted. Then Boulware and Deser showed that the solution with $+ ve$  branch is unstable and the solution is asymptotically an $AdS$ Schwarzschild solution with negative gravitational mass, indicating the instability. The solution (\ref{sol1}) with $- ve$ branch is stable and the solution is asymptotically a Schwarzschild solution. Therefore the  $+ ve$  branch  is of less physical interest  \cite{Nojiri:2001aj,Nojiri:1999ji}
}

This describes a $5D$ $AdS$ regular black hole  for EGB gravity coupled 
with NLED. The resulting black hole is  characterized by parameters like   
$M$,   $k$ and $\alpha$. In the limit,   $\alpha\to 0$ and $k=0$, the negative branch of 
solution (\ref{eqnf}) corresponds to the $5D$  Schwarzschild-Tangherlini black hole. 
However,  in  the limit  $\alpha\rightarrow 0$, the solution (\ref{eqnf}) corresponds  
to the regular Schwarzschild black hole  in $5D$ $AdS$ \cite{hc,Balart:2014cga}
\begin{equation}
 f(r)=\left(1-\frac{2Me^{-k/r^2}}{r^2}- \frac{\Lambda r^2}{3}\right).
\label{eqn.gb}
\end{equation}
Here, we remark that the exponential factor present in the solution removes the 
curvature singularity. The given metric  (\ref{eqnf}) can also be considered as the EGB 
black hole coupled to NLED. It can be checked that  solution  (\ref{eqnf}) matches with 
the Boulware-Deser black hole provided the  mass $(M)$ must be replaced with $M(r)$:
\begin{equation}
M(r)=\frac{\sigma(r)}{\sigma_{\infty}}M,
\end{equation}
where   $\sigma(r)=e^{-k/r^2}$ is the probability distribution function satisfying $
\sigma(r)\geq 0$ and  $\sigma'(r)< 0$ for $r\geq 0$. Also,  $\sigma(r)/r\to 0$ for 
$r\to 0$ and $\sigma_\infty$ refers to is the probability distribution function when 
$r\rightarrow\infty$.  Asymptotically 
($(r>>k)$),  the metric   (\ref{eqnf})  corresponds  to the charged $AdS$ EGB black 
hole  \cite{Wiltshare88}  
\begin{equation}
f(r)=1+\frac{r^2}{4\alpha}\left(1\pm\sqrt{1+\frac{8\alpha M}{r^4}-\frac{8\alpha q^2}
{r^6}+ \frac{8\Lambda\alpha}{3}}\,\right).
\end{equation}
Henceforth, we end up with a new solution describing an exact regular  EGB  black hole  
 coupled with nonlinear matter fields in $AdS$ space.  This $AdS$ solution, 
 characterized by    the parameter  $M$ and $k$, extends the Wiltshire charged 
 EGB black hole  \cite{Wiltshare88} to $AdS$ space. 

 The horizon of the black hole is described by the following condition:
\begin{equation}
1+\frac{r^2}{4\alpha}\left(1\pm \sqrt{1+\frac{8M\alpha}{r^{4}}e^{-\frac{k}{r^{2}}}+ 
\frac{8\Lambda\alpha}{3}}\,\right)=0.\label{11}
\end{equation} 
\begin{figure*}[ht]
\begin{tabular}{c c c c}
\includegraphics[width=.5\linewidth]{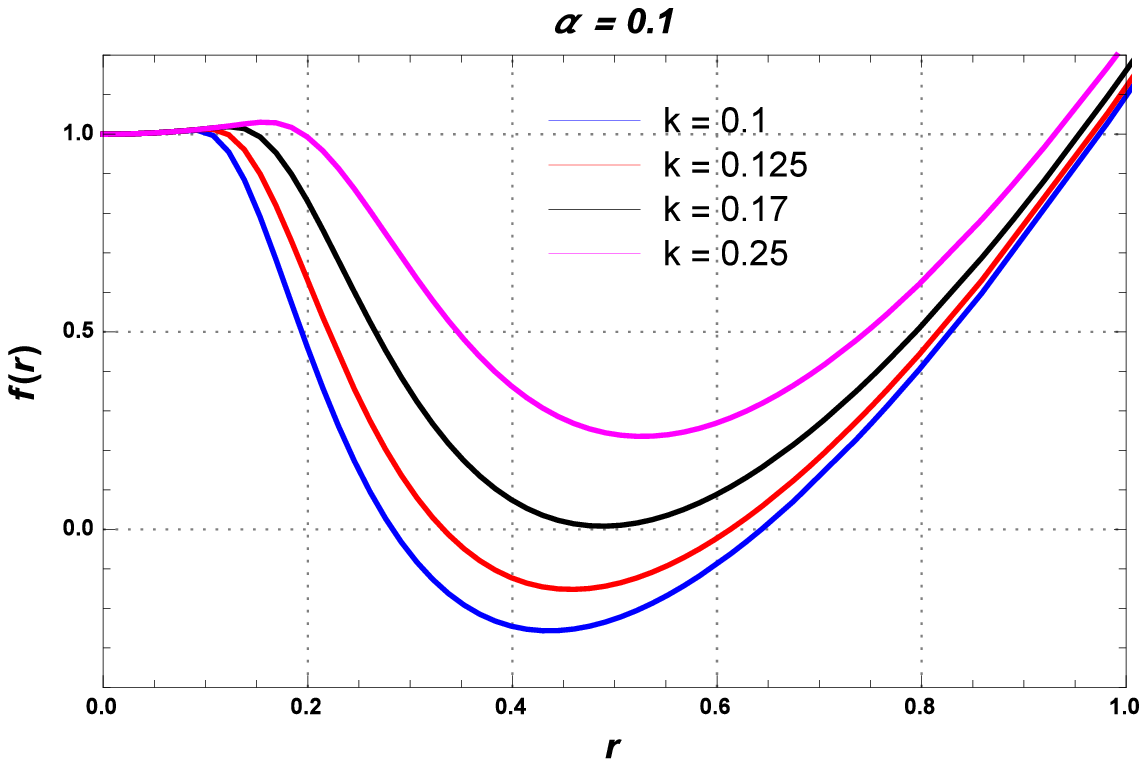}
\includegraphics[width=.5\linewidth]{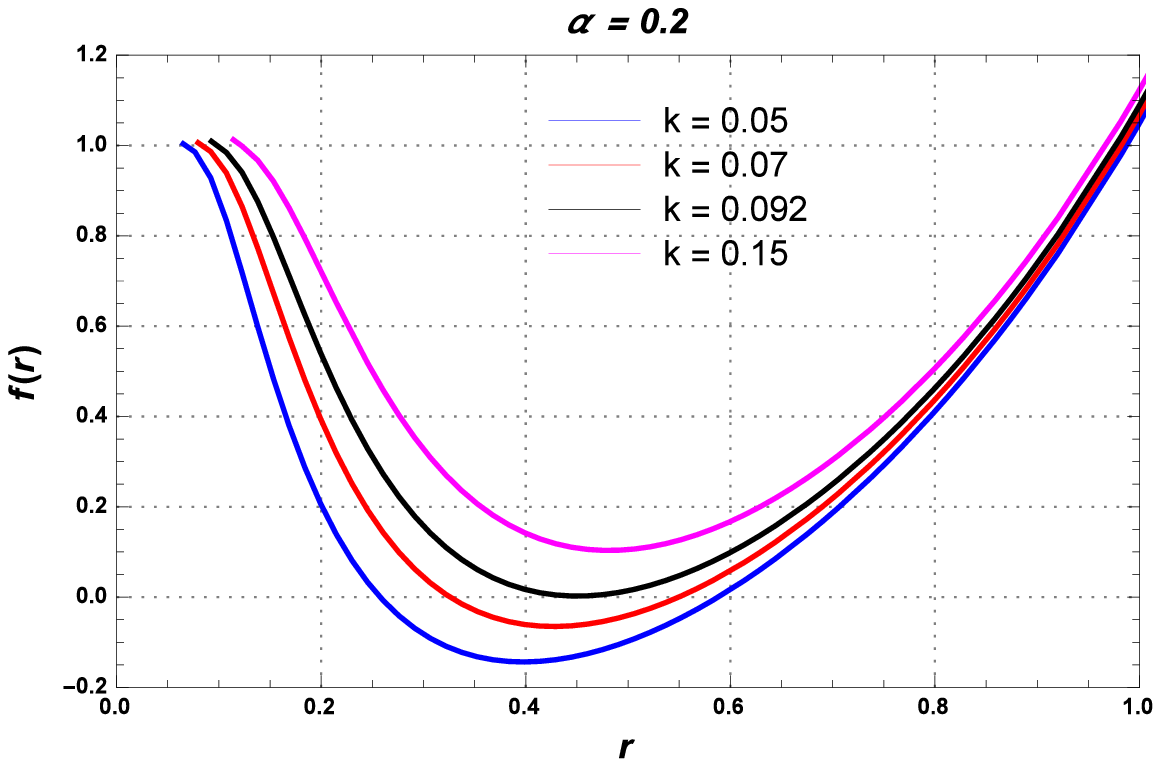}
\end{tabular}
\caption{ The  metric function $f(r)$ versus  $r$ with different value of deviation parameter $k$ for $\alpha=0.1$ (left panel) and $\alpha=0.2$ (right panel) with fixed  $M$ and $l$.}
\label{fig:1}
\end{figure*}
The plot for $f(r)$ versus $r$ is depicted in FIG. \ref{fig:1}. 
Eq. (\ref{11}) is a complex expression that complicates the analysis of the horizon structure analytically. Henceforth, we prefer numerical analysis of the horizon condition by varying the deviation parameter $k$.  The horizon condition  $ f(r)=0$ will find two real roots, namely,  $r_{+}$ and $r_{-}$ that correspond to the event and Cauchy horizon, respectively.
The numerical values of $r_{-}$ and $r_{+}$ for different $\alpha$ and $k$ are tabulated in Table \ref{tab:temp1}.
\begin{table}[h]
		\begin{center}
			\begin{tabular}{| l r l l| l r l l  |}
				\hline		\hline			 
				\multicolumn{1}{|c}{ }&\multicolumn{1}{c}{ $\alpha=0.1$}&\multicolumn{1}{c}{}&\multicolumn{1}{c|}{  }&\multicolumn{1}{c}{}&\multicolumn{1}{c}{ $\alpha=0.2$}&\multicolumn{1}{c}{}&\multicolumn{1}{c|}{}\\
				\hline
				\multicolumn{1}{|c}{  {$k$}} & \multicolumn{1}{c}{ $r_-$ } & \multicolumn{1}{c}{ $r_+$ }& \multicolumn{1}{c|}{$\delta$}&\multicolumn{1}{c}{ {$k$}}& \multicolumn{1}{c}{$r_-$} &\multicolumn{1}{c}{$r_+$} &\multicolumn{1}{c|}{ $\delta$}   \\
				\hline
				\,\,\,0.1\,\,& \,\,0.283\,\, &\,\, 0.643\,\,& \,\,0.360\,\,&\,\,  0.1 &\,\, 0.258\,\,&\,\,0.596\,\,&\,\,0.338\,\,
				\\
				\,\,\,0.125\,\,& \,\,0.329\,\, &\,\, 0.612\,\,& \,\,0.283\,\,&\,\,  0.07 &\,\, 0.329\,\,&\,\,0.558\,\,&\,\,0.229\,\,
				\\
				\,\, 0.17\,\, & \,\,0.483\,\, &\,\, 0.483\,\,& \,\,0\,\,&\,\, 0.092 &\,\,0.447\,\,&\,\,0.447\,\,&\,\,0\,\,
				\\
				\hline\hline	
		 			\end{tabular}
		\end{center}
		\caption{Cauchy  horizon ($ r_{-}$). event  horizon ($ r_{+}$), and $\delta= r_+- r_-$ for  the $5D$  $AdS$ EGB Bardeen black hole with ${\alpha}=0.1$  and ${ \alpha}=0.2$ with fixed   $M$ and $\Lambda$.}
		\label{tab:temp1}
	\end{table}
The horizons can also be discussed in terms of the deviation parameter $k$. Now, it is possible to compute the value of $k$ which satisfies the horizon condition that admits two real roots for $r$.  
From the table, it is evident that there exists a critical horizon  $r_{c}=r_{\pm}=0.483$ and critical deviation parameter $k_c=0.17$  for $\alpha=0.1$. However, a critical horizon  $r_{c}=r_{\pm}=0.447$ and critical deviation parameter $k_c=0.092$ 
exist for $\alpha=0.2$.  These signify extremal regular $AdS$  black holes.
Moreover, $k<k_{c}$ for  $\alpha=0.1$ and $k>k_{c}$ for  $\alpha=0.2$, the two different  horizons ($r_{\pm}$) signify the non-extremal black hole.  We find that the size of the black hole decreases with an increase in the value of  $\alpha$.

	\section{Thermodynamics}\label{sec3}

\begin{figure*}[ht]
\begin{tabular}{c c c c}
\includegraphics[width=.5\linewidth]{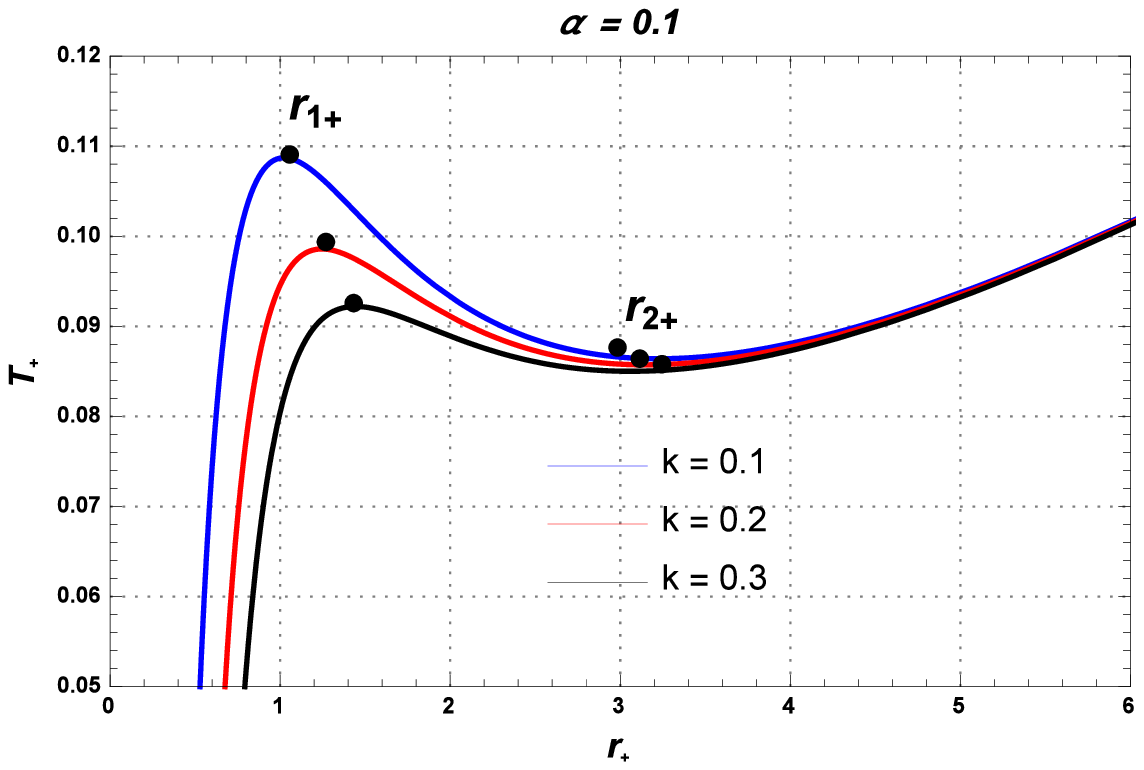}
\includegraphics[width=.5\linewidth]{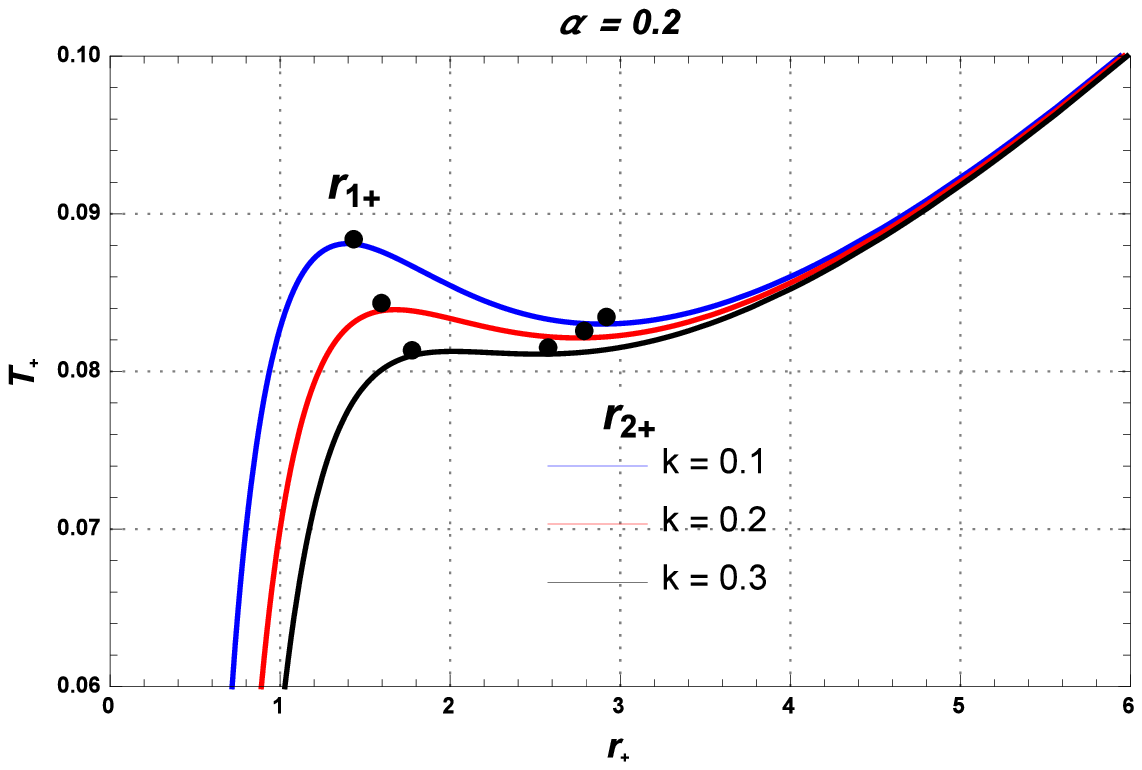}
\end{tabular}
\caption{Temperature $T_+$ versus $r_+$ for distinct value of deviation parameter $k$ with $\alpha=0.1$ (left panel) and $\alpha=0.2$ (right panel)  with fixed   $M=1$ and $l=10$. }
\label{fig:2}
\end{figure*}

Now, we can study the thermodynamic properties of the obtained black hole solution in terms of horizon radius, which are described by the horizon mass $(M_+)$, deviation parameter $(k)$, and the cosmological constant $\Lambda$. The horizon mass and Hawking temperature  ($T_+$) is calculated by
\begin{eqnarray}
&&M_{+} =e^{k/r_+^2}\left(\frac{r_+^4}{l^2}+ (r_+^2 + 2\alpha)\right),\label{M}\\
&&T_+ =\left.\frac{1}{4\pi}\frac{\partial}{\partial r}\sqrt{-g^{rr}g_{tt}}\right|_{r=r_+} =\frac{r_+^4 -k(r_+^2 + 2\alpha)}{2\pi r_+^3 (r^2 + 4\alpha)}+\frac{2r_+^6-k r_+^4  }{2l^2 \pi r_+^3 (r_+^2 + 4\alpha)}\label{t}.
\end{eqnarray}
The Hawking temperature is characterized by   $k$,   $\alpha$, and   $\Lambda$. The temperature of this regular  $AdS$ black hole is plotted in FIG. \ref{fig:2}. 
From the figure, we see that the effect of the deviation parameter is more significant for small black holes. 
As the value of $k$ increases, the peak of the temperature decreases and occurs at larger $r_+$ as well. The temperature of the $AdS$ regular black hole also decreases with an increase in $\alpha$ and shifts toward the large value of the horizon radius.

Being a thermal system, the black hole follows the first law of thermodynamics 
  given by
\begin{equation}
dM_{+} = T_{+}dS_{+}+\phi de,
\end{equation}
where $S_+$ refers to the entropy of the black hole.
For the given values of $M_+$ and $T_+$, the first law of thermodynamics leads to the
following expression for the entropy:
\begin{equation}
    S_+=\frac{4\pi r_+^3}{3}\left[\frac{(2k+r_+^2+12\alpha)e^{k/r_+^2}}{r_+^2}-\frac{2\sqrt{\pi k}\,(4k+6\alpha)}{r_+^3}\text{erf}\left(\frac{\sqrt{k}}{r_+}\, \right)\right],
\end{equation}
where erf is the error function. Here, we note that the entropy matches with the one
calculated in Ref. \cite{Ghosh:2018bxg} without the cosmological constant. The deformed   entropy from the area-law occurs due to  the presence of deviation parameters and   GB parameter.

 We know that    entropy of the regular black hole does not follow the area law \cite{Ansoldi:2008jw, Bronnikov:2000vy}
because the black hole mass is included in the source term. Ma {\it et al} \cite{ma14} proposed the corrected form of first law  black hole thermodynamics for regular black holes which modifies with the extra factor. The modified first law is   \cite{ma14,Singh:2022xgi,M2}
\begin{equation}
C_{M}dM=T_+\,dS +  \phi \,de,
\label{mod}
\end{equation}
 where  $C(M,r_+)$ is 
\begin{equation}
C(M,r_+)=1+4\pi \int_{r_+}^{\infty}r^2\frac{\partial T^0_0}{\partial M} dr=2 e^{-k/r_+^2}.
\end{equation}
For this value of $C(M,r_+)$     and the obtained black hole solution follows the area law.

\begin{figure*}[ht]
\begin{tabular}{c c c c}
\includegraphics[width=.5\linewidth]{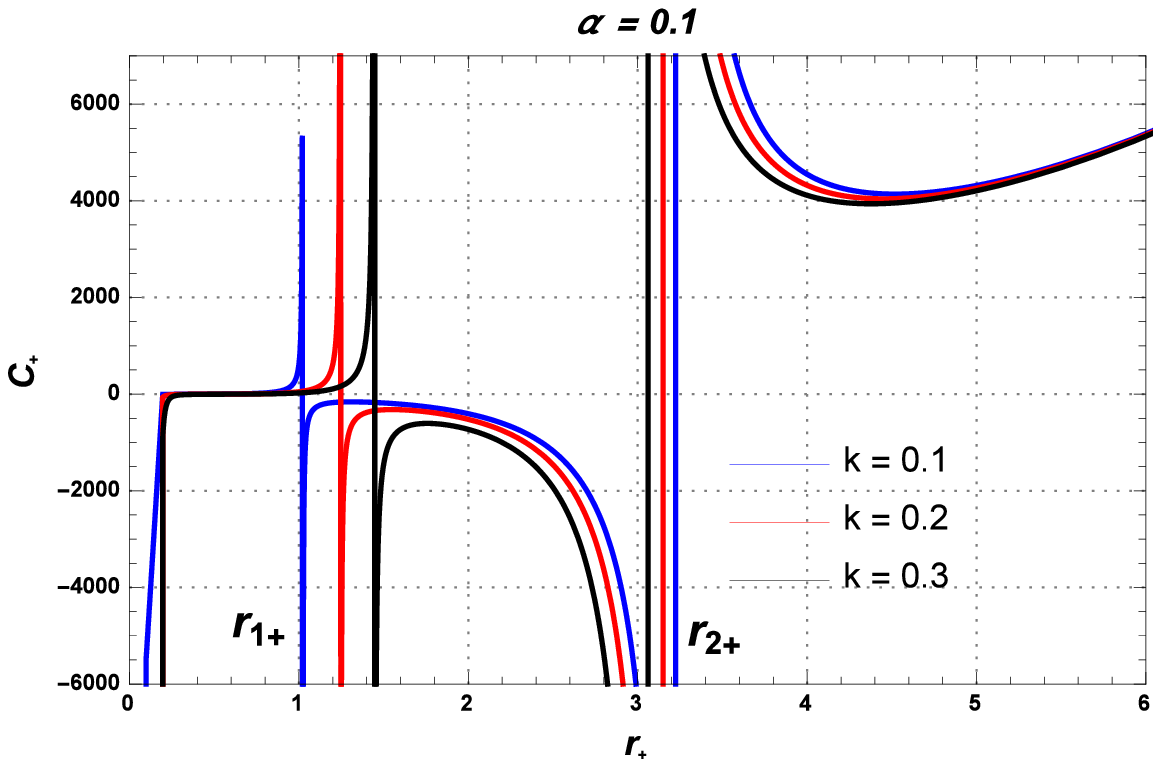}
\includegraphics [width=.5\linewidth]{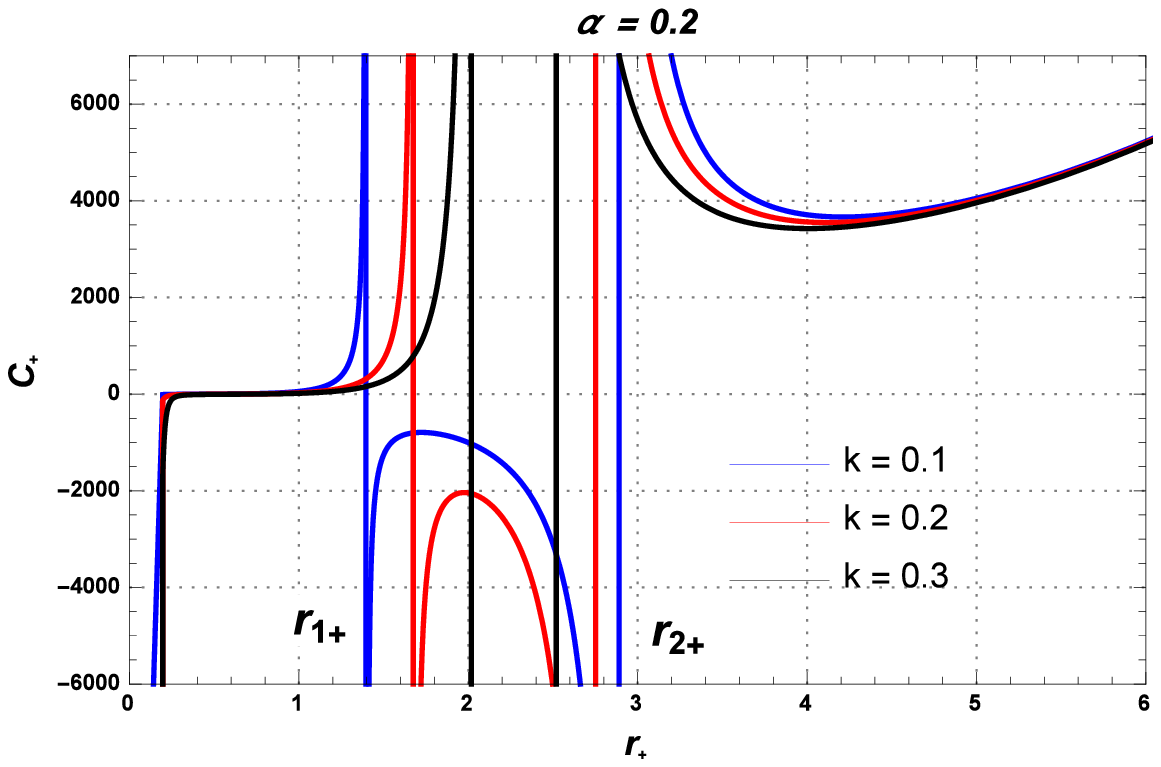} 
\end{tabular}
\caption{The   heat capacity $C_+$  versus $r_+$ with different value of deviation parameter $k$ for $\alpha=0.1$ (left panel) and $\alpha=0.2$ (right panel) with fixed value of $M=1$ and $l=10$. }
\label{sh1}
\end{figure*}

The thermodynamic stability  of the given black hole can be explained by the nature of the heat capacity as the positive and negative signatures of heat capacity justify
the stable and unstable state of the black hole, respectively.
The heat capacity for the black hole solution can be defined as
\begin{eqnarray}\label{SH}
C_+&=&  \frac{\partial{M_+}}{\partial{T_+}}.
\end{eqnarray}
For the given   mass (\ref{M}) and temperature (\ref{t}), 
 the expression of the heat capacity reads
 \begin{eqnarray}
C_+=\frac{4e^\frac{k}{r_+^2}\pi r(r_+^2 +4\alpha)^2 (r_+^4(l^2 + 2r_+^2)-k(r_+^4 +l^2(r_+^2 +2\alpha)))}{2r_+^6(r_+^2 +12\alpha)-l^2 (r_+^6 -4r_+^4\alpha) +k(r_+^6 -4r_+^4\alpha +l^2(3r_+^4 +14r_+^2 \alpha +24\alpha^{2}))}.
\label{SH1}
\end{eqnarray}
\noindent From this expression,  it is cumbersome to identify the signature and  behavior of heat capacity. Hence, we plot the heat capacity as depicted in diagram \ref{sh1} for different values of deviation parameter $k$.
To study the nature of the heat capacity, we  plotted them for the various values of $k$ and $\alpha$.
From the  FIG. \ref{sh1}, we find that there exist  double phase transitions.  Firstly, a phase transition occurs from a small   stable black hole to a large  unstable black hole and, secondly, from a smaller unstable black hole to a larger stable black hole. For the fixed value of $\alpha$ the radii $r_{1+}$ increases and $r_{2+}$ decreases  with $k$.

\begin{figure*}[ht]
\begin{tabular}{c c c c}
 \includegraphics[width=.5\linewidth]{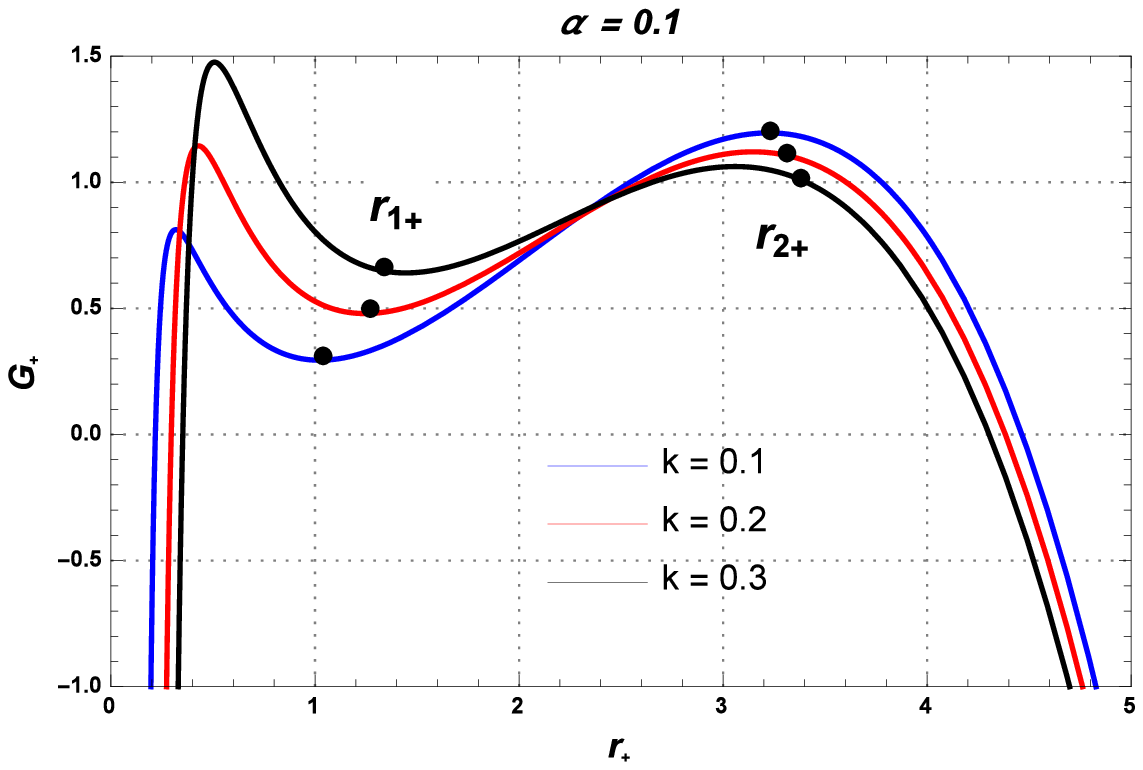}
\includegraphics[width=.5\linewidth]{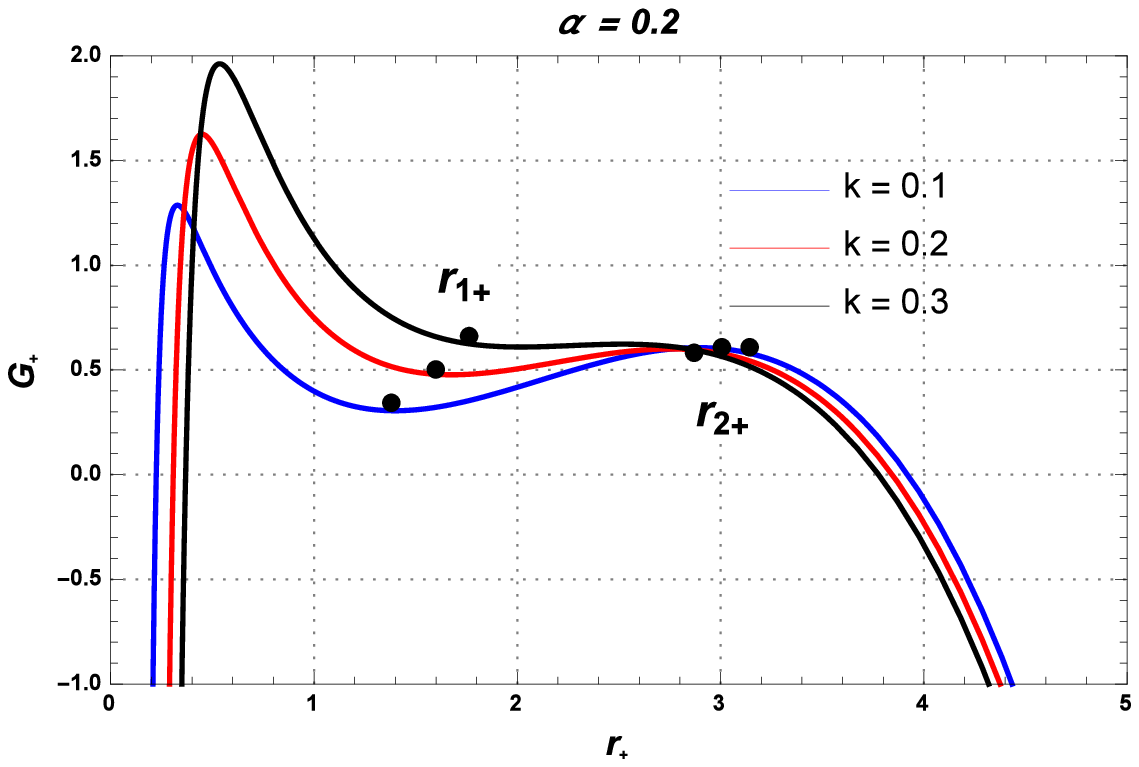}
\end{tabular}
\caption{The Gibbs free energy $G_+$ versus $r_+$ with different value of deviation parameter $k$ for $\alpha=0.1$ and $\alpha=0.2$ with fixed value of $M=1$ and $l=10$. }
\label{sh2}
\end{figure*}

Gibbs free energy also plays an important role in order to discuss
the  (global) stability of the black hole. The Gibbs free energy can be calculated from the standard definition:  $G_+=M_+-T_+S_+$.
This yields
\begin{eqnarray}
G_+&=&e^{k/r_+^2}\left(\frac{r_+^4}{l^2}+ (r_+^2 + 2\alpha)\right)-\frac{2}{3}\left(\frac{r_+^4(l^2+2r_+^2)-k(r_+^4+l^2(r_+^2+2\alpha))}{ l^2(r_+^2+4\alpha)}\right)\nonumber\\&\times & \left[\frac{(2k+r_+^2+12\alpha)e^{k/r_+^2}}{r_+^2}-\frac{2\sqrt{\pi k}\,(4k+6\alpha)}{r_+^3}\text{erf}\left(\frac{\sqrt{k}}{r_+}\, \right)\right].
\end{eqnarray}
The stability can also be explained from the Gibbs free energy plot as depicted in FIG. \ref{sh2}.
From the plot, we observe that   the free energy exhibits  local minima ($r_{1+}$) and local maxima ($r_{2+}$) for distinct values of $k$ with fixed value of $\alpha$.
 For $r > r_{1+}$, the free energy is an increasing function of   $r_+$ and remains positive    and attains the local maximum value at $ r_{2+}$. After  $r= r_{2+}$, the slope of Gibbs free energy turns negative and, therefore,  the theory  provides the natural Hawking-Page phase transition.
The various numerical values are tabulated in the TABLE \ref{tab2}. 
 Here, we can also see that the Gibbs free energy ($G_+$) has a  minimum and a   maximum locally regarding the extremal points of the temperature where the heat capacity diverges.   
	\begin{table}[h]
		\begin{center}
			\begin{tabular}{|l l l l l| l l l l l|}
				\hline
				\hline
				\multicolumn{1}{|c}{ }&\multicolumn{1}{c}{  }&\multicolumn{1}{c}{$\alpha=0.1$ }&\multicolumn{1}{c}{ }&\multicolumn{1}{c|}{ }&\multicolumn{1}{c}{ }&\multicolumn{1}{c}{ }&\multicolumn{1}{c}{$\alpha=0.2$} &\multicolumn{1}{c}{}&\multicolumn{1}{c|}{}\\
						\hline
				\,\,\,$r_{1+}$\,\,& \,\,$r_{2+}$\,\, &\,\, $T_+$\,\,& \,\,$C_+$\,\,&\,\,\,\,$G_+$&\,\,\,$r_{1+}$\,\,& \,\,$r_{2+}$\,\, &\,\, $T_+$\,\,& \,\,$C_+$\,\,&\,\,\,\,$G_+$
				\\
					\hline	 	
			 \,\,1.012  \,\,& \,\,3.218\,\, &\,\,  Maximum\,\,& \,\,Diverge\,\,&\,\,Loc. Max.\,\, &\,\,1.405  \,\,&\,\,2.881\,\,&\,\,Minimum\,\,& \,\,Diverge\,\,&\,\,Loc. Min.\,\,
				\\ 	
				 \,\,1.251\,\,& \,\,3.148\,\, &\,\, Maximum\,\,& \,\,Diverge\,\,&\,\,Loc. Max.\,\,&\,\,1.714\,\,&\,\,2.754\,\,&\,\,Minimum\,\,& \,\,Diverge\,\,&\,\,Loc. Min.\,\,
				\\
				\,\,1.447\,\, & \,\,3.063\,\, &\,\, Maximum\,\,& \,\,Diverge\,\,&\,\,Loc. Max.\,\,&\,\,2.09\,\,&\,\,2.515\,\,&\,\,Minimum\,\,& \,\,Diverge\,\,&\,\,Loc. Min.\,\,
				\\
				 \hline
				\hline
			\end{tabular}
		\end{center}
		\caption{The numerical values of the local maxima and minima to characterize the nature of $T_+$, $C_+$, and $G_+$.}
		\label{tab2}
	\end{table}


\section{Van der Waals Fluid}\label{sec4}
The aim of this section is to consider the resulting black hole as a Van der Waals fluid and calculate the $P-v$ criticality. As we know, the negative cosmological constant induces  a thermodynamic pressure  (i.e. $\Lambda=-8\pi P_+$  with $G = \hbar = c = 1$) in the extended thermodynamics.  The   thermodynamic volume $V$ plays the role of  conjugate to pressure and can be interpreted as
 the change in the mass under the variations in the $\Lambda$ having fixed  horizon area. The mass $M$ is then understood as an
enthalpy.
     
     The temperature $T_{+}$ in tandem to the above identifications of thermodynamic pressure and conjugate volume lead the following
equation of state:
Using the  and volume $V$, we  obtain  the following equations of state:
\begin{eqnarray}
P_{+}=\frac{1}{4\pi r^4(2r_+^2-k)}\left[6\pi r_+^3 T(r_+^2+4\alpha)+ 3(kr_+^2-r_+^4+2k\alpha) \right],\,\,\,\,   v=2r_+,
\end{eqnarray}
where the $v$ is  a specific volume. 

\begin{figure*}[ht]
\begin{tabular}{c c c c}
\includegraphics[width=.5\linewidth]{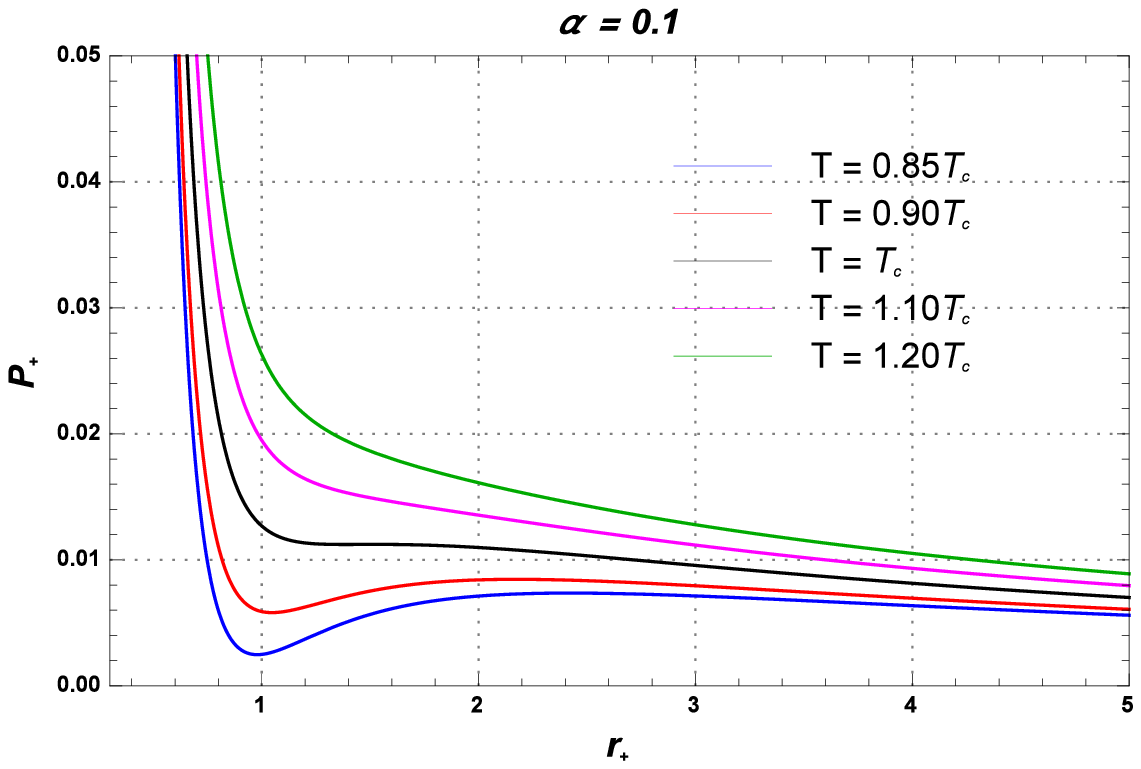}
\includegraphics [width=.5\linewidth]{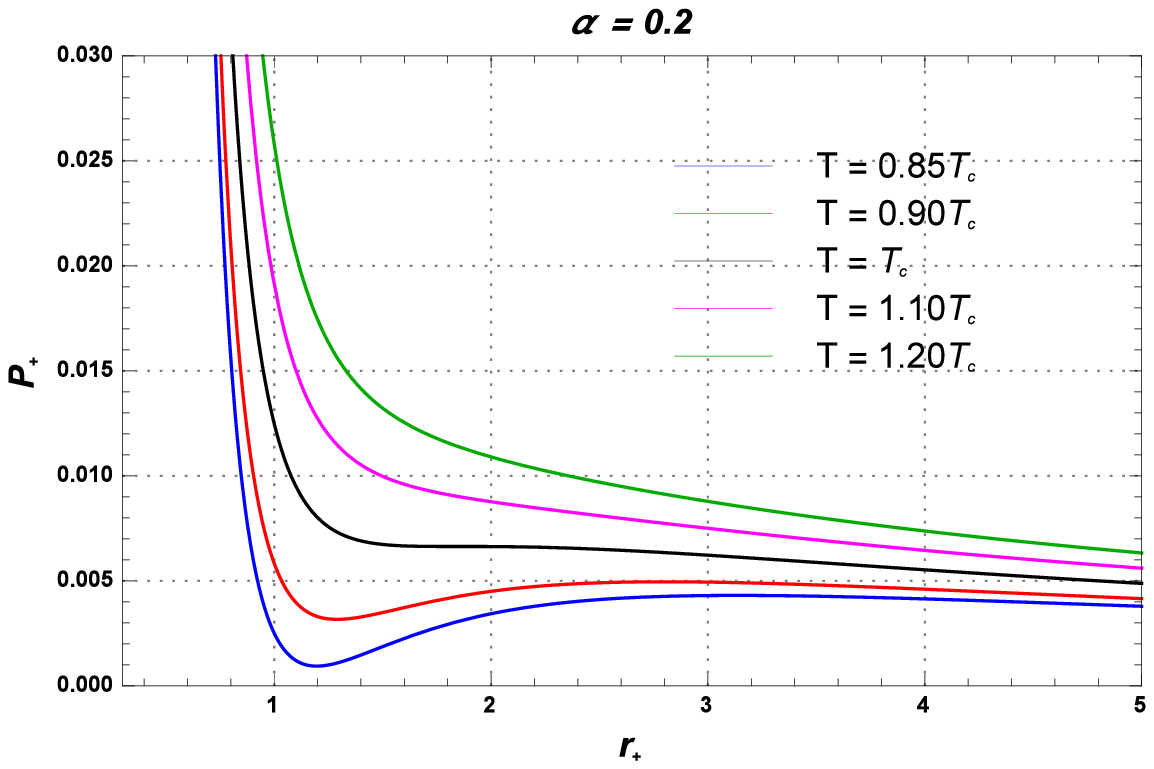}\\
\includegraphics[width=.5\linewidth]{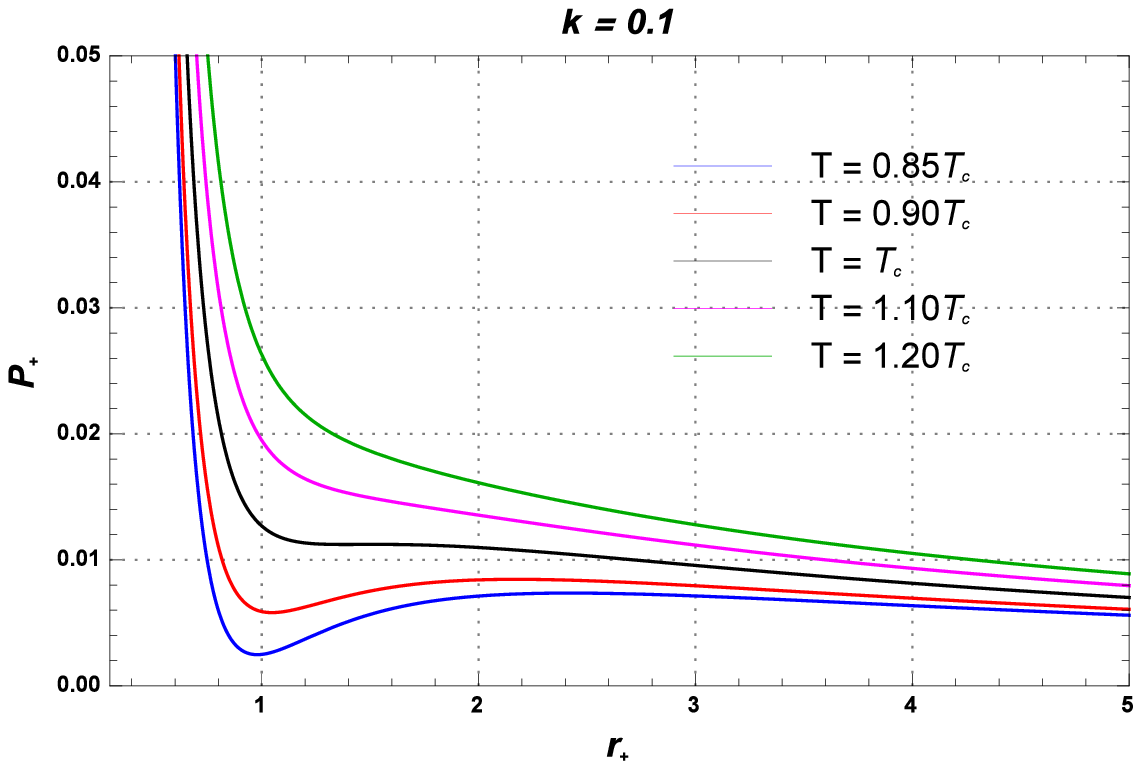}
\includegraphics[width=.5\linewidth]{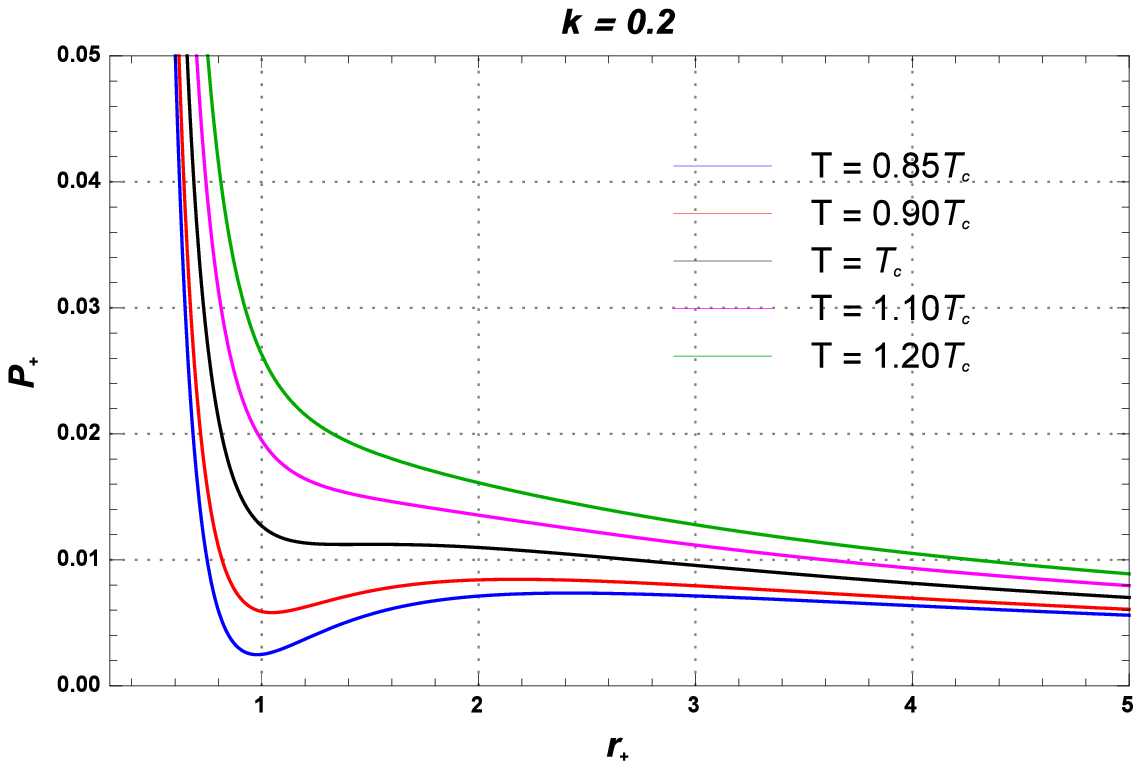}
\end{tabular}
\caption{  $P_+$  versus $r_+$ with different value of deviation parameter $k$ for $\alpha=0.1$ and $\alpha=0.2$ at critical temperature $T_c$. }
\label{sh0}
\end{figure*}

The critical points appear at
isotherms  $T_c$  where pressure  has an inflection point at  $P_c$  and  $v_c$ satisfying conditions [10,14]
\begin{eqnarray}
\left(\frac{\partial P_{+}}{\partial r_+} \right)_{T_+}=0,\qquad
\left(\frac{\partial^2 P_{+}}{\partial r^2_+}\right)_{T_+}=0.
\end{eqnarray}
The critical radius can be calculated numerically from the relation
\begin{eqnarray}
\frac{ 2r^8-24r^6\alpha-k^2(3r^4+16r^2\alpha-48\alpha^2)-12k(r^6+9r^4\alpha+36r^2\alpha^2) }{4\pi r^6(kr^2+2r^4-4k\alpha+24r^2\alpha)(k-2r^2)}=0.\label{pvv}
\end{eqnarray}
The fluid behavior of the regular $AdS$ EGB black holes   can be seen from the $P_+-r_+$ diagram in FIG.  \ref{sh0}.

The  Eq. (\ref{pvv}) can not be solved analytically and, therefore, the critical radius $r_c$, critical pressure $P_c$ and the critical temperature  $T_c$ are obtained numerically. The numerical values are presented in  TABLES \ref{tr10} and    \ref{tr12} for different values of $\alpha$ and $k$.
\begin{table}[ht]
 \begin{center}
 \begin{tabular}{ |l | l   | l   | l   | l|   }
\hline
            \hline
  \multicolumn{1}{|c|}{ $k$} &\multicolumn{1}{c}{$r_c$}  &\multicolumn{1}{|c|}{$T_c$}  &\multicolumn{1}{c|}{$P_c$} &\multicolumn{1}{c|}{${P_c\,r_c}/{T_c}$}\\
            \hline
            \,\,\,\,\,0.1~~ &~~1.453~~ & ~~0.123~~ & ~~0.0225~~ &  ~~0.2173~~      \\
            \,\,\,\,\,0.2~~ &~~1.689~~  & ~~0.110~~ & ~~0.0176~~ &  ~~0.2425~~    \\
            \,\,\,\,\,0.3~~ &~~1.884~~ & ~~0.100~~ & ~~0.0146~~ &    ~~0.2518~~  \\
            \,\,\,\,\,0.4~~ &~~2.055~~  & ~~0.093~~ & ~~0.0125~~ &    ~~0.2558~~   \\
            \,\,\,\,\,0.5~~ &~~2.209~~  & ~~0.087~~ & ~~0.0107~~ &    ~~0.2579~~    \\
            \hline 
\hline
        \end{tabular}
        \caption{The  critical temperature $T_c$, critical pressure $P_c$ and $P_c\,r_c/T_c$ corresponding different value of $k=0.1$ with fixed value of $\alpha=0.1$ .}
\label{tr10}
    \end{center}
\end{table}
\begin{table}
 \begin{center}
 \begin{tabular}{ |l | l   | l   | l   |  l|   }
\hline
            \hline
  \multicolumn{1}{|c|}{ $\alpha$} &\multicolumn{1}{c|}{$r_c$}  &\multicolumn{1}{c|}{$T_c$}  &\multicolumn{1}{c|}{$P_c$} &\multicolumn{1}{|c|}{$P_c\,r_c/T_c$}\\
            \hline
            \,\,\,\,\,0.1 ~~  &~~1.453~~  & ~~0.123~~ & ~~0.0225~~ & ~~0.2173~~ \\            
            \,\,\,\,\,0.2~~ &~~1.8434~~ & ~~0.0936~~ & ~~0.0138~~ &  ~~0.2617~~      \\
            \,\,\,\,\,0.3~~ &~~2.154~~  & ~~0.0785~~ & ~~0.0094~~ &  ~~0.3216~~    \\
            \,\,\,\,\,0.4~~ &~~2.423~~ & ~~0.0690~~ & ~~0.0073~~ &    ~~0.4069~  \\
            \,\,\,\,\,0.5~~ &~~2.662~~  & ~~0.0623~~ & ~~0.0060~~ &    ~~0.5381~~   \\
            \hline 
\hline
        \end{tabular}
        \caption{The   critical temperature $T_c$, critical pressure $P_c$ and $P_c\,r_c/T_c$ corresponding to the different value of   $\alpha=0.1$ with fixed value of $k=0.1$ .}
\label{tr12}
    \end{center}
\end{table}
  It can be seen that the  critical  radius $r_c$  increases with the increase in the parameters $k$ and $\alpha$, however, the critical pressure $P_c$ and temperature  $T_c$  decrease  with increase in $k$ and $\alpha$. Incidentally, the universal ratio $P_cr_c/T_c$   increases with the parameters $k$ and  $\alpha$. It is worth mentioning that  the critical radius  increases with decrease in the critical pressure   and critical temperature.

\section{QNMs in Ekilon Limit}\label{sec5}
 QNMs  are usually predicts 
the stability of the given black holes   perturbed by an external
field or    black hole geometry.  QNMs
also provide the information regarding gravitational waves. 
In   QNM can be discussed by  studying the motion of photon in the vicinity  of the black hole solution (\ref{eqnf}). The   photon motion limited to equatorial plane $(\theta=\pi/2)$ is described by the following Lagrangian:
\begin{equation}
{\cal {L}}=-g_{tt}{\dot t}^2 +g_{rr}{\dot r}^2+g_{\theta\theta}{\dot \theta}^2+g_{\phi\phi} {\dot \phi}^2+g_{\psi\psi} {\dot \psi}^2  ,
\label{lag1}
\end{equation}
where dot denotes the derivative with respect to affine parameter.
The corresponding Hamiltonian  is given by
 \begin{equation}
{\cal{H}}=\frac{1}{2}g^{ij}p_ip_j=0,
 \end{equation}
 and the generalized momenta are given by
\begin{eqnarray}
&&p_t=\frac{\partial {\cal H}}{\partial {\dot t}}=\text{constant}\equiv E, \qquad 
 p_r=\frac{\partial {\cal H}}{\partial {\dot r}}=g_{rr}{\dot r}, \qquad  p_{\theta}
 =\frac{\partial {\cal H}}{\partial {\dot \theta}}=g_{\theta\theta}{\dot\theta},\nonumber \\
&&p_{\phi}=\frac{\partial {\cal H}}{\partial {\dot \phi}}=\text{constant} \equiv -L, \qquad  p_{\psi}=\frac{\partial {\cal H}}{\partial {\dot \psi}}=g_{\psi\psi}{\dot\psi}.
\end{eqnarray}
Here, $ E$ and $L$ refer to the energy and the angular momentum
per unit rest mass of the test particle, respectively.
The equations of motion  associated with the photon in the Hamiltonian formalism are given by
\begin{eqnarray} 
&&{\dot t}=\frac{\partial {\cal H}}{\partial p_t}=-\frac{p_t}{g_{tt}},\quad {\dot r}=\frac{\partial {\cal H}}{\partial p_r}=-\frac{p_r}{g_{rr}}, \quad{\dot \theta}=\frac{\partial {\cal H}}{\partial p_{\theta}}=\frac{p_{\theta}}{g_{\theta\theta}}, \nonumber\\
&&{\dot \phi}=\frac{\partial {\cal H}}{\partial p_{\phi}}=\frac{p_{\phi}}{g_{\phi\phi}},\quad{\dot \psi}=\frac{\partial {\cal H}}{\partial p_{\psi}}=\frac{p_{\theta}}{g_{\psi\psi}}.
\end{eqnarray}
since the above Hamiltonian does not depend on the coordinates   $t$,   $\phi$ and $\psi$. So, the null geodesics equation is written by  
\begin{equation}
    {\dot r}^2+V_{eff}(r)=0, \qquad \text{with} \qquad V_{eff}=f(r)\left(\frac{L^2}{r^2}-\frac{E^2}{f(r)}\right).
\end{equation}
For a circular null geodesics, the effective potential  must  necessarily hold the following conditions:
\begin{equation}
V_{eff}=0, \qquad \text{and}\qquad \frac{\partial V_{eff}}{\partial r}=0.
\label{pot}
\end{equation}
These conditions describe the radius of the photon sphere.
These conditions    lead  to the  
 equation of the photon radius as
\begin{equation}
kM-2Mr^2+e^{k/r^2}r^2\sqrt{1+\frac{8M\alpha e^{k/r^2}}{r^4}+\frac{8\Lambda\alpha}{3r^2}}=0.
\label{pr1}
\end{equation} 
This equation can not be solved analytically, so we can calculate the photon radius $r_p$, numerically. The numerical values are presented in TABLE \ref{tr1}. From this TABLE, we can see that the photon radius increases along with  increasing deviation parameter and GB coupling.

\begin{table}[ht]
 \begin{center}
 \begin{tabular}{ |l | l   | l   | l   |  l |  }
\hline
            \hline
  \multicolumn{1}{|c|}{ $$} &\multicolumn{1}{c}{$$}  &\multicolumn{1}{c}{$r_p$}  &\multicolumn{1}{c }{ } &\multicolumn{1}{c|}{}\\
            \hline
  \multicolumn{1}{|c|}{ $\alpha$} &\multicolumn{1}{c|}{$k=0.1$}  &\multicolumn{1}{c|}{$k=0.2$}  &\multicolumn{1}{c|}{$k=0.3$} &\multicolumn{1}{|c|}{$k=0.4$}\\
            \hline
            \,\,\,\,\,0.1 ~~  &~~1.4381~~  & ~~1.4574~~ & ~~1.4726~~ & ~~1.4840~~ \\            
            \,\,\,\,\,0.2~~ &~~1.4503~~ & ~~1.4783~~ & ~~1.5021~~ &  ~~1.5169~~      \\
            \,\,\,\,\,0.3~~ &~~1.4728~~  & ~~1.5154~~ & ~~1.5479~~ &  ~~1.5727~~    \\
            \,\,\,\,\,0.4~~ &~~1.5298~~ & ~~1.6023~~ & ~~1.6817~~ &    ~~1.6954~  \\
            \hline 
\hline
        \end{tabular}
        \caption{The numerical values of photon radius corresponding to the GB coulpling parameter  and devation parameter  with     $M=1$, where $l=1$ and $n=0$.}
\label{tr1}
    \end{center}
\end{table}

\begin{table}[ht]
 \begin{center}
 \begin{tabular}{ |l | l   | l   | l   |  l |  }
\hline
            \hline
  \multicolumn{1}{|c|}{ } &\multicolumn{1}{c}{}  &\multicolumn{1}{c}{$r_p$}  &\multicolumn{1}{c}{} &\multicolumn{1}{c|}{}\\
            \hline
  \multicolumn{1}{|c|}{ $k$} &\multicolumn{1}{c|}{$\alpha=0.1$}  &\multicolumn{1}{c|}{$\alpha=0.2$}  &\multicolumn{1}{c|}{$\alpha=0.3$} &\multicolumn{1}{|c|}{$\alpha=0.4$}\\
            \hline
            \,\,\,\,\,0.1 ~~  &~~1.438~~  & ~~1.450~~ & ~~1.472~~ & ~~1.529~~ \\            
            \,\,\,\,\,0.2~~ &~~1.457~~ & ~~1.478~~ & ~~1.515~~ &  ~~1.603~~      \\
            \,\,\,\,\,0.3~~ &~~1.472~~  & ~~1.500~~ & ~~1.547~~ &  ~~1.654~~    \\
            \,\,\,\,\,0.4~~ &~~1.484~~ & ~~1.516~~ & ~~1.572~~ &    ~~1.695~  \\
            \,\,\,\,\,0.5~~ &~~1.491~~  & ~~1.528~~ & ~~1.591~~ &    ~~1.727~~   \\
            \hline 
\hline
        \end{tabular}
        \caption{Values of photon radius corresponding to the GB coulpling parameter $(\alpha)$ and devation parameter $(k)$ with  $M=1$, where $l=1$ and $n=0$.}
\label{tr012}
    \end{center}
\end{table}

The QNMs frequency $\omega$  in the eikonal limit  can be estimated by the virtue  of the photon sphere as follows
\begin{equation}
\omega=l\Omega-i\left(n+\frac{1}{2}\right)|\Lambda|,    
\end{equation}
where $n$ is the overtone number and $l$ is the angular quantum number of perturbation. However, $\Omega$ is the angular velocity  and $\Lambda$ is the  Lyapunov exponent   of the photon sphere with following expressions:
\begin{eqnarray}
 \Omega=\frac{\sqrt{f(r_p)}}{r_p} \qquad \text{and}\qquad \Lambda=\frac{\sqrt{f(r_p)(2f{r_p}-r^2_pf''(r_p))}}{\sqrt{2} r_p}.
\end{eqnarray}
Here $r_p$  denotes radius of photon sphere (called as photon radius). 

\begin{figure*}[ht]
\begin{tabular}{c c c c}
\includegraphics[width=.5\linewidth]{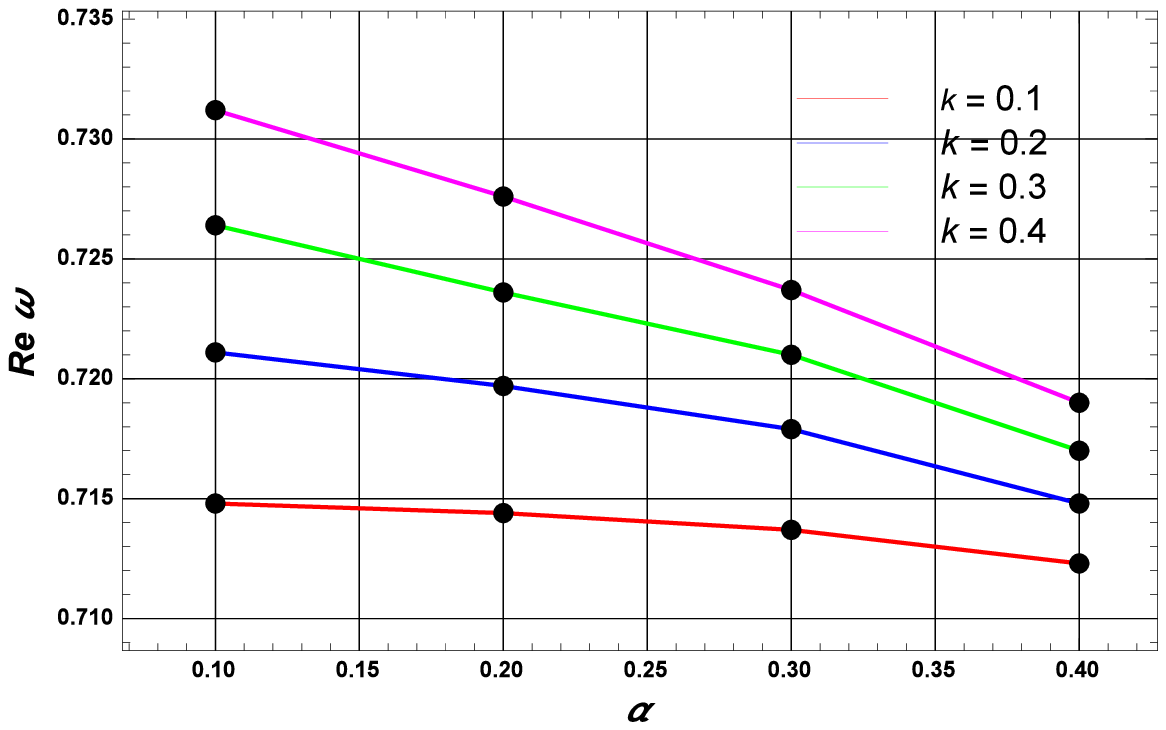}
\includegraphics [width=.5\linewidth]{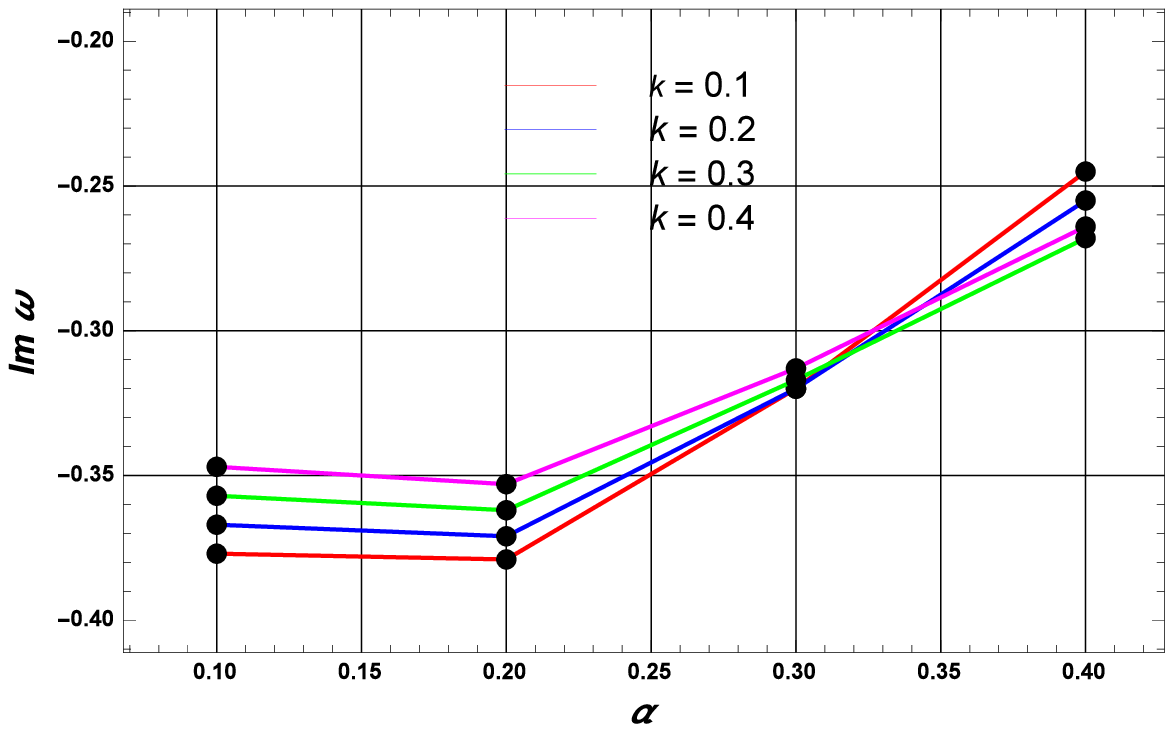}
\end{tabular}
\caption{The plot of real part (left panel)  and imaginary part (right panel) of QNMs versus GB parameter with different   $k$  with  fixed  $M$ and $l$. }
\label{sh}
\end{figure*}

\begin{figure*}[ht]
\begin{tabular}{c c c c}
\includegraphics[width=.5\linewidth]{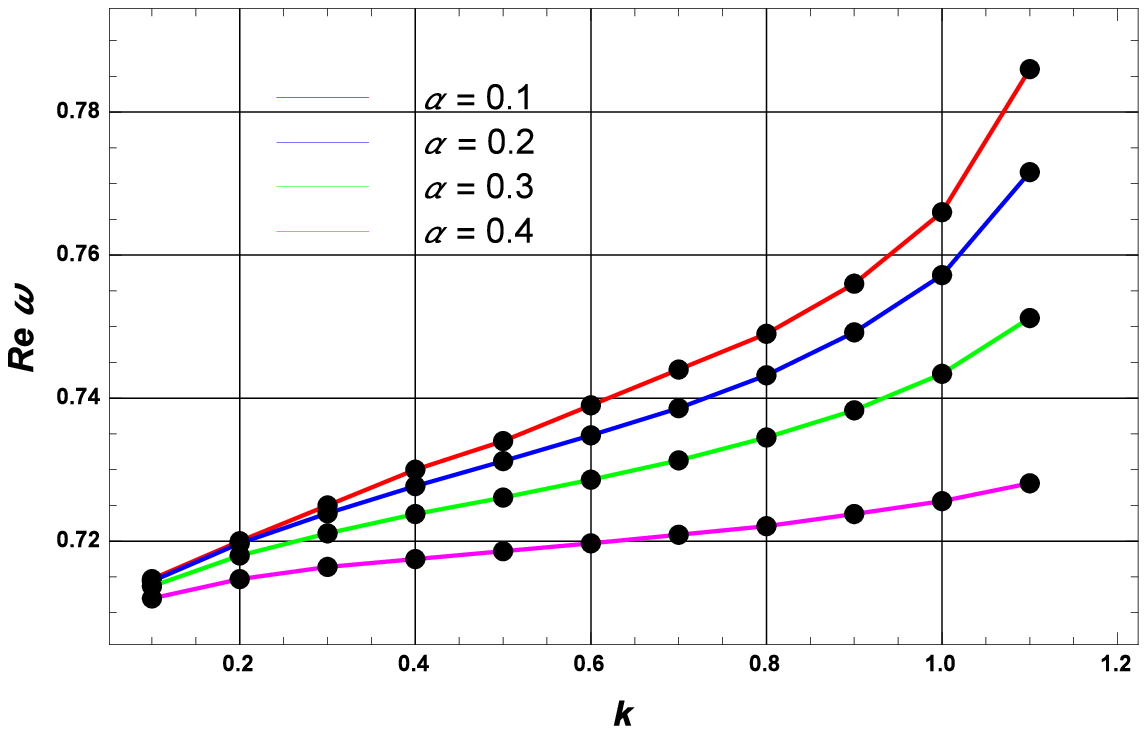}
\includegraphics [width=.5\linewidth]{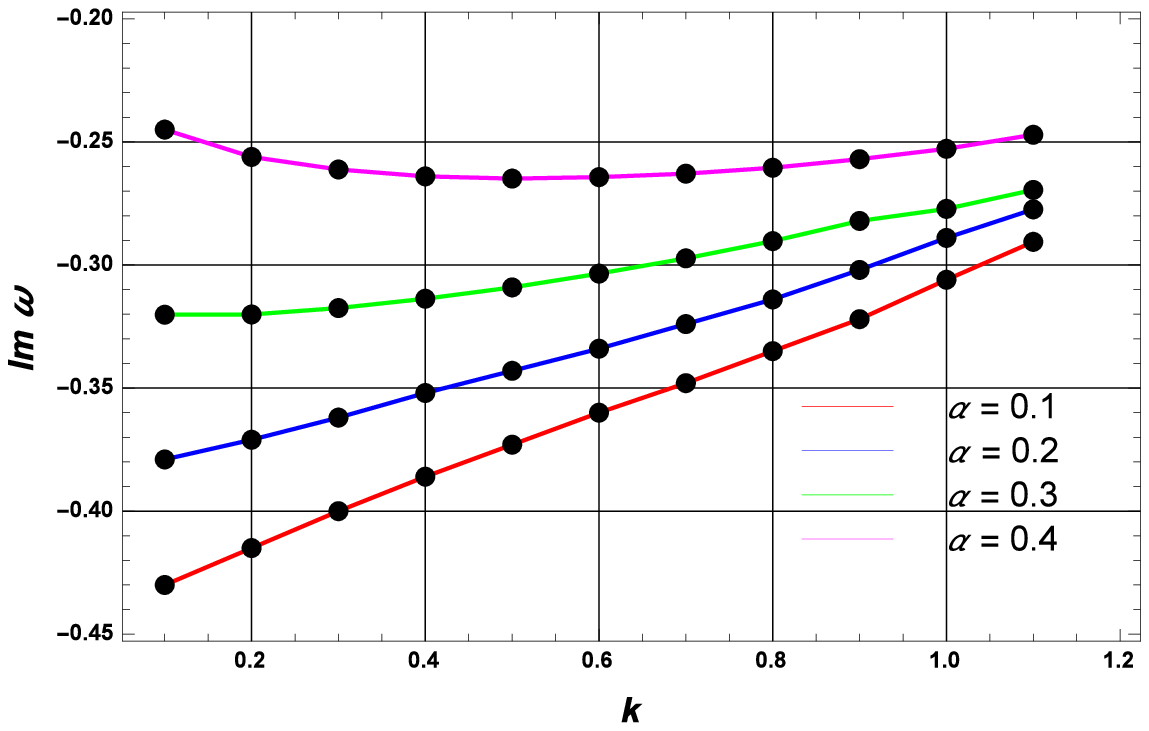}
\end{tabular}
\caption{The plot of real part (left panel)  and imaginary part (right panel) of QNMs versus deviation parameter with different  $k$  with   fixed $M$ and $l$. }
\label{sh11}
\end{figure*}

The real   and imaginary parts  of the QNMs of black hole solution (\ref{eqnf})   for different values of deviation parameter and GB  parameter are depicted in the FIG.  \ref{sh} and FIG. \ref{sh11}. 
These diagrams  help us  to  investigate the effects of the black hole parameters on the QNMs. Here, we  see that the real part of the QNMs is a decreasing function of the GB parameter. However, real part of the QNMs increases with deviation parameter. 
On the other hand,  the imaginary part of the QNMs with respect to the GB parameter   first decreases very slowly (almost constant) and then increases sharply. Also, the imaginary part of the QNMs increases with deviation parameter (more significantly for large $\alpha$).

The signature of the imaginary part of the QNMs characterizes the stability of black hole.  Im $\omega<0$ corresponds to stable modes of black hole and Im $\omega>0$ corresponds to unstable modes. The imaginary part of the QNMs  for the obtained  black hole solution (\ref{eqnf}) is negative (See Fig. \ref{sh} and \ref{sh11}). This confirms  that the modes of the obtained  black hole solution (\ref{eqnf}) are stable.

We list the numerical values of QNMs frequency corresponding various values of parameters in TABLES \ref{tr13} and \ref{tr14}.

\begin{table}[ht]
 \begin{center}
 \begin{tabular}{| l | l   | l   | l   |  l|   }
\hline
            \hline
  \multicolumn{1}{|c|}{ } &\multicolumn{1}{c|}{$k=1$}  &\multicolumn{1}{c|}{$k=2$}  &\multicolumn{1}{c|}{$k=3$} &\multicolumn{1}{|c|}{$k=4$}\\
  \hline
    \multicolumn{1}{|c|}{ $\alpha$} &\multicolumn{1}{c|}{$\omega$=Re \,$\omega$ $-$ Im \,$\omega$}  &\multicolumn{1}{c|}{$\omega$=Re \,$\omega$ $-$ Im \,$\omega$}  &\multicolumn{1}{c|}{$\omega$ =Re \,$\omega$ $-$ Im \,$\omega$} &\multicolumn{1}{|c|}{$\omega$=Re \,$\omega$ $-$ Im \,$\omega$}\\
            \hline
            \,\,\,\,\,0.1 ~~  &~~0.71489 - 0.37765 $i$~~  & ~~0.721135 - 0.36758 $i$~~ & ~~0.726456 - 0.35734 $i$~~ & ~~0.73125 - 0.34703 $i$~~ \\            
            \,\,\,\,\,0.2~~ &~~0.71440 - 0.37984 $i$~~ & ~~0.719702 - 0.37125 $i$~~ & ~~0.723687 - 0.36266 $i$~~ &  ~~0.72764 - 0.35300 $i$~~      \\
            \,\,\,\,\,0.3~~ &~~0.71376 - 0.32043 $i$~~  & ~~0.717979 - 0.32024 $i$~~ & ~~0.721104 - 0.31778 $i$~~ &  ~~0.72375 - 0.31388 $i$~~    \\
            \,\,\,\,\,0.4~~ &~~0.71238 - 0.24545 $i$~~ & ~~0.714816 - 0.25595 $i$~~ & ~~0.714815 - 0.26844 $i$~~ &    ~~0.71756 - 0.26416 $i$~  \\
            \hline 
\hline
        \end{tabular}
        \caption{The numerical values of QNMs corresponding to the GB coulpling parameter $(\alpha)$ and devation parameter $(k)$ with    $M=1$, where $l=1$ and $n=0$.}
\label{tr13}
    \end{center}
\end{table}

\begin{table}[ht]
 \begin{center}
 \begin{tabular}{ |l | l   | l   | l   |  l|   }
\hline
            \hline
  \multicolumn{1}{|c|}{ } &\multicolumn{1}{c|}{$\alpha=0.1$}  &\multicolumn{1}{c|}{$\alpha=0.2$}  &\multicolumn{1}{c|}{$\alpha=0.3$} &\multicolumn{1}{|c|}{$\alpha=0.4$}\\
  \hline
    \multicolumn{1}{|c|}{ $k$ }  &\multicolumn{1}{c|}{$\omega$=Re \,$\omega$ $-$ Im \,$\omega$}  &\multicolumn{1}{c|}{$\omega$=Re \,$\omega$ $-$ Im \,$\omega$}  &\multicolumn{1}{c|}{$\omega$=Re \,$\omega$ $-$ Im \,$\omega$} &\multicolumn{1}{|c|}{$\omega$=Re \,$\omega$ $-$ Im \,$\omega$}\\
            \hline
            \,\,\,\,\,0.1 ~~  &~~0.7147 -0.4307 $i$~~  & ~~0.7144 -0.3797 $i$~~ & ~~0.7137 -0.3202 $i$~~ & ~~0.71241 -0.245 $i$~~ \\            
            \,\,\,\,\,0.2~~ &~~0.7208 -0.4151 $i$~~ & ~~0.7197 -0.3712 $i$~~ & ~~0.7180 -0.3201 $i$~~ &  ~~0.7147 -0.2561 $i$~~      \\
            \,\,\,\,\,0.3~~ &~~0.7258 -0.4004 $i$~~  & ~~0.7239 -0.3621 $i$~~ & ~~0.7211 -0.3175 $i$~~ &  ~~0.7164 -0.2612 $i$~~    \\
            \,\,\,\,\,0.4~~ &~~0.7303 -0.3866 $i$~~ & ~~0.7277 -0.3528 $i$~~ & ~~0.7238 -0.3137 $i$~~ &    ~~0.7175 -0.2640 $i$~  \\
            \,\,\,\,\,0.5~~ &~~0.7347 -0.3733 $i$~~  & ~~0.7312 -0.3435 $i$~~ & ~~0.7261 -0.3091 $i$~~ &    ~~0.7186 -0.2649 $i$~~   \\
            \,\,\,\,\,0.6~~ &~~ 0.7391 -0.3607 $i$~~ & ~~0.7348 -0.3341 $i$~~ & ~~0.7286 -0.3035 $i$~~ &    ~~0.7197 -0.2643 $i$~~   \\
            \,\,\,\,\,0.7~~ &~~0.7440 -0.3482 $i$~~  & ~~0.7386 -0.3245 $i$~~ & ~~0.7313 -0.2973 $i$~~ &    ~~0.7209 -0.2629 $i$~~   \\
               \,\,\,\,\,0.8~~ &~~0.7496 -0.3356 $i$~~  & ~~0.7432 -0.3143 $i$~~ & ~~0.7345 -0.2903 $i$~~ &    ~~0.7221 -0.2605  $i$~~   \\
              \,\,\,\,\,0.9~~ &~~0.7568 -0.3220 $i$~~  & ~~0.7492 -0.3028 $i$~~ & ~~0.7383 -0.2821 $i$~~ &    ~~0.7238 -0.2570 $i$~~   \\
              \,\,\,\,\,1.0~~ &~~0.7669 -0.3061 $i$~~  & ~~0.7572 -0.2891 $i$~~ & ~~0.7434 -0.2772 $i$~~ &    ~~0.7256 -0.2528 $i$~~   \\
              \,\,\,\,\,1.1~~ &~~0.7861 -0.2806 $i$~~  & ~~0.7716 -0.2674 $i$~~ & ~~0.7512 -0.2695 $i$~~ &    ~~0.7281 -0.2471 $i$~~   \\
            \hline 
\hline
        \end{tabular}
        \caption{The numerical values of QNMs corresponding to the GB coulpling parameter $(\alpha)$ and devation parameter $(k)$ with   $M=1$, where $l=1$ and $n=0$.}
\label{tr14}
    \end{center}
\end{table}

\section{Results and Conclusion}\label{sec6}

In this work, we have considered a EGB 
gravity coupled to the NLED in  $5D$ $AdS$ spacetime and constructed a new regular  black hole  solution in $AdS$ spacetime. The obtained solution is a generalized version of 
$5D$ Schwarzschild-Tangherlini black hole, $5D$ $AdS$ regular Schwarzschild black hole
and Boulware-Deser black hole. We have found that the black hole solution
exhibits    two  horizons, namely,  the Cauchy and event horizon.
There exist different critical horizons corresponding to different GB parameter that
characterize the  extremal/non-extremal nature of black holes.  The size of the black holes decrease with the increasing GB parameter.

Furthermore, we have studied the thermodynamics of the resulting solution by deriving horizon mass, Hawking temperature and entropy of the black hole. We have found that
the black hole satisfies the modified first law of thermodynamics. The stabilities of black hole are discussed by estimating both the heat capacity and Gibbs free energy.
The diagrams confirmed  that there exist
double phase transitions, one from small stable black hole to large unstable black hole
and  other  from smaller unstable black hole to larger stable black hole.
The    Gibbs free energy analysis confirms the existence of (local) minimum and   maximum associated  to the extremal points of the
Hawking temperature. The fluid nature of black hole is also studied. We have observed  that the critical values depend  on GB coupling parameter and deviation parameter considerably.
For instance, the critical radius is an increasing function of the GB coupling parameter and deviation parameter. In contrast, the critical pressure   and critical temperature are  decreasing  function of the GB coupling parameter and deviation parameter. 
 
It is worth discussing  QNMs for the $5D$ $AdS$ regular EGB black hole coupled with NLED as QNMs may provide the information regarding gravitational waves. For this purpose, we studied the motion of photon in the vicinity of the black hole solution.
The effects of GB coupling and deviation parameters on the real and imaginary parts of the QNMs are also discussed.  It will be interesting to establish a the correspondence between the QNMs in the eikonal limit and the shadow radii  for such black hole solution of the $5D$ EGB gravity coupled to the NLED in $AdS$ space.

\begin{acknowledgements}  
This research was funded by the Science Committee of the Ministry of Science
and Higher Education of the Republic of Kazakhstan (Grant No. AP09058240).
One of us (DVS)   thanks UGC for the start up grant (Grant No.: 30-600/2021(BSR)/1630). 
\end{acknowledgements}

\section*{Data Availability Statement and  Competing Interests} 
Data sharing not applicable to this article as no datasets were generated or analysed during the current study. The authors declare no competing interests.

\end{document}